\documentclass[fleqn,10pt]{wlscirep}
\usepackage[T1]{fontenc}
\usepackage{graphicx}
\usepackage{epsfig}
\usepackage{epstopdf}
\usepackage{dcolumn}
\usepackage{bm}
\usepackage{mathrsfs}
\usepackage{svg} 
\usepackage[mathlines]{lineno}
\usepackage[utf8]{inputenc}
\usepackage{textcomp}
\linenumbers\relax 

\linenumbers

\title{Orbital Fulde-Ferrell-Larkin-Ovchinnikov state in 2H-NbS\textsubscript{2} flakes}

\author[1,2,+]{Xinming Zhao}
\author[2,3,+]{Guoliang Guo}
\author[1,2,4,+,*]{Chengyu Yan}
\author[5]{Noah F.Q. Yuan}
\author[6]{Chuanwen Zhao}
\author[1,2]{Huai Guan}
\author[1,2]{Changshuai Lan}
\author[1,2]{Yihang Li}
\author[2,3,4,* ]{Xin Liu}
\author[1,2,4,*]{Shun Wang}

\affil[1]{Ministry of Education Key Laboratory of Fundamental Physical Quantities Measurement and Hubei Key Laboratory of Gravitation and Quantum Physics, National Precise Gravity Measurement Facility , Huazhong University of Science and Technology, Wuhan 430074, People’s Republic of China}
\affil[2]{School of Physics, Huazhong University of Science and Technology, Wuhan 430074, People’s Republic of China}
\affil[3]{Wuhan National High Magnetic Field Center,Huazhong University of Science and Technology, Wuhan 430074, People’s Republic of China}
\affil[4]{Institute for Quantum Science and Engineering, Huazhong University of Science and Technology, Wuhan 430074, People’s Republic of China}
\affil[5]{Tsung-Dao Lee Institute, Shanghai Jiao Tong University, Shanghai 201210, China and School of Physics and Astronomy, Shanghai Jiao Tong University, Shanghai 200240, People’s Republic of China}
\affil[6]{State Key Laboratory of Magnetic Resonance and Atomic and Molecular Physics, National Center for Magnetic Resonance in Wuhan, Wuhan Institute of Physics and Mathematics, Innovation Academy for Precision Measurement Science and Technology, Chinese Academy of Sciences, Wuhan 430071, Hubei, People’s Republic of China}
\affil[+]{X. Zhao, G. Guo  and C. Yan contribute equally to the work.}
\affil[*]{corresponding author(s): Chengyu Yan(chengyu$\_$yan@hust.edu.cn), Xin Liu(phyliuxin@hust.edu.cn),Shun Wang(shun@hust.edu.cn)}
\nolinenumbers
\begin{abstract}
Symmetry breaking in a layered superconductor with Ising spin-orbit coupling has offered an opportunity to realize unconventional superconductivity. To be more specific, orbital Fulde-Ferrell-Larkin-Ovchinnikov (FFLO) state, exhibiting layer-dependent finite-momentum pairing, may emerge in transition metal dichalcogenides materials (TMDC) in the presence of an in-plane magnetic field. Orbital FFLO state can be more robust against the magnetic field than the conventional superconducting state with zero-momentum pairing. This feature renders its potential in field-resilient superconducting functionality. Although, orbital FFLO state has been reported in NbSe\textsubscript{2} and MoS\textsubscript{2}, it is not yet clear if orbital FFLO state can be extended to other TMDC superconductor. Here, we report the observation of orbital FFLO state in 2H-NbS\textsubscript{2} flakes and its dependence on the thickness of flake. We conclude that the relatively weak interlayer coupling is instrumental in stabilizing orbital FFLO state at higher temperature with respect to the critical temperature and lower magnetic field with respect to paramagnetic limit in NbS\textsubscript{2} in comparison to its NbSe\textsubscript{2} counterpart. 
\end{abstract}

\begin{document}                              
\flushbottom
\maketitle

\thispagestyle{empty}


\section*{Introduction} 

Superconductivity with Cooper pairs carrying finite pairing momentum has recently stimulated opportunities to realize emerging functionality operating under conditions not feasible for superconductivity with zero-momentum pairing. For instance, finite-momentum pairing can result in superconducting diode effect exhibiting non-reciprocal supercurrent\cite{HNC2023,XL2023,BBF2022,BSS2024}. It has been known for decades that finite-momentum pairing can be routinely induced by breaking time-reversal symmetry via a magnetic field, manifested as Fulde-Ferrell-Larkin-Ovchinnikov (FFLO) state\cite{YF2022,G1960,FF1964,KSL2020,BMC2003,SK2019,CK2024}. However, the realization of FFLO state is far from easy\cite{SK2019,KMS2022,GG1966} owing to the requirement of large magnetic field and low temperature. Fortunately, the stringent condition of activating finite-momentum pairing can be relaxed by simultaneously breaking both inversion and time-reversal symmetries\cite{XL2023,LT2012,NL2021,WZY2023}. This renders Rashba-superconductors\cite{NL2021,ZGZ2014,BG2002} and Ising-superconductors\cite{XL2023,WZY2023,ZL2022}, with intrinsic inversion-symmetry breaking, suitable candidates for the task. 

\begin{figure}[t]
    \centering
\includegraphics[width=0.9\textwidth]{F0.jpg}
\caption{\textbf{A brief summary of unconventional superconductivity in TMDC materials}. In the few-layer limit, Ising superconductivity has been widely observed\cite{DSG2018,XWZ2016,JOI2015}, it is characterized by the gigantic enhancement of the in-plane upper critical field $B_{c2,||}$. Usually, $B_{c2,||}$ can exceed Pauli limit $B_{p}$ at a relatively high temperature with respect to critical temperature $T_{c}$. Ising superconductivity is with zero pairing momentum. In the bulk limit, Zeeman FFLO driven by Zeeman effect of the in-plane magnetic field is observed in 2H- NbS\textsubscript{2}\cite{CLN2021,CCC2022}. A noticeable enhancement in $B_{c2,||}$ occurs at a relatively low temperature when the field approaches $B_{p}$. Zeeman FFLO is with finite pairing momentum. Note that there are also reports on Ising superconductivity in the bulk \cite{CTK2024,FHS2024}. In the flakes with intermediate thickness, orbital FFLO has been reported in NbSe\textsubscript{2}\cite{WZY2023} and Li-intercalated MoS\textsubscript{2}\cite{ZDK2023}. The enhancement of $B_{c2,||}$ kicks in at intermediate temperature and field. Orbital FFLO is also with finite pairing momentum. Our work will focus on the flakes with intermediate thickness. }
\label{fig1}
\end{figure}

It is particularly interesting to focus on an exotic FFLO state in transition metal dichalcogenides (TMDC) superconductors with Ising spin-orbit coupling (SOC), referred to as orbital FFLO state\cite{WZY2023,Y2023}. Orbital FFLO state has only been recently firmly realized in two TMDC materials, namely Li-intercalated  MoS\textsubscript{2}\cite{ZDK2023} and NbSe\textsubscript{2}\cite{WZY2023}. There are two reasons making orbital FFLO unique: First, the role of the in-plane magnetic field is dramatically different in orbital FFLO state compared to that in standard FFLO state (named as Zeeman FFLO) or FFLO state in Rashba-superconductors (labeled as Rashba FFLO). The in-plane magnetic field primarily induces Zeeman splitting in Zeeman FFLO and Rashba FFLO\cite{YF2022,G1960,FF1964,NL2021,ZGZ2014}, though the impact of Zeeman effect can be reduced  at most by a factor of $\sqrt{2}$ in the case of Rashba FFLO. On the contrary, Zeeman effect of in-plane magnetic field is suppressed by orders of magnitude in the presence of Ising SOC, instead, the in-plane magnetic field triggers a layer-dependent shift in momentum via interlayer orbital effect\cite{XL2023,ZL2022,WZY2023,Y2023,L2017}. A much smaller field is require to stabilize orbital FFLO state compared to its Zeeman FFLO counterpart. Second, the orbital FFLO state can potentially participate in intriguing interplay with other unconventional superconducting states in an Ising-superconductor. The unconventional superconductivity in TMDC materials is briefly summarized in Fig.\ref{fig1}. Ising superconductivity primarily dominates in few-layer limit\cite{DSG2018,XWZ2016,JOI2015} along with a few reports on Ising superconductivity in the bulk \cite{CTK2024,FHS2024}, orbital FFLO state is reported in flakes with intermediate thickness\cite{WZY2023}, meanwhile there is incipient evidence of Zeeman FFLO in bulk\cite{CLN2021,CCC2022}. It is highly interesting to determine the phase boundary between these state as a function of flake thickness.  

\begin{figure}[h]
    \centering
\includegraphics[width=0.68\textwidth]{F1.jpg}
\caption{\textbf{Basic characterization of the 2H-NbS\textsubscript{2} flakes}. \textbf{a.} Typical temperature dependence of a NbS\textsubscript{2} flake with a thickness of 19 nm. \textbf{b.} Zero-field critical temperature as a function of thickness. The experimental data is well captured by Eq.\ref{eq1}. Inset shows the optical image of a typical device.}
\label{fig2}
\end{figure}

Since the study on oebital FFLO state is in its infancy, it is urgent to verify the existence of orbital FFLO in other materials and summarize the common features. In this work, we report the observation of orbital FFLO state in high-quality 2H-NbS\textsubscript{2} flakes, the only material in the TMDC family that is free of charge density wave. It is noticed that orbital FFLO state in NbS\textsubscript{2} qualitatively resembles that in MoS\textsubscript{2}\cite{ZDK2023} and NbSe\textsubscript{2}\cite{WZY2023}, but also exhibiting quantitative difference owning to the difference in the strength of spin-orbit coupling and interlayer coupling. More interestingly, assembling our work that focuses on flakes of intermediate thickness along with Ising superconductivity widely observed in TMDC few-layers\cite{XWZ2016,JOI2015,DSG2018} and Zeeman FFLO that so far only has been detected in NbS\textsubscript{2} bulk in the entire TMDC family \cite{CLN2021,CCC2022}, it highlights that NbS\textsubscript{2} can potentially be a model system to study the thickness-driven interplay between  Ising superconductivity, orbital FFLO state and Zeeman FFLO state.

\begin{figure*}[t]
    \centering
\includegraphics[width=0.68\textwidth]{F2.jpg}
\caption{\textbf{Evidence of orbital FFLO state in a 19 nm flake.} \textbf{a.} Evolution of upper critical field against temperature. It is clear that in-plane upper critical field $B_{c2,||}$ can be nicely fitted by orbital FFLO model. Detail of the model can be found in Supplemental Material Material Sec.1 and Sec.4\cite{SM}. \textbf{b\&c.} $B_{c2,||}$ as a function of polar angle $\theta$ measured at \(T = 5.7 K > T^*\)  and \(T = 4 K < T^*\). The in-plane orientation is defined as $\theta=0^\circ$. The angle has been carefully calibrated as shown in Fig.S1(c). }
\label{fig3}
\end{figure*}

\section*{Results}

Observation of orbital FFLO state puts an unprecedented requirement on sample quality to mitigate spurious effects such as surface oxidization\cite{LIAN2017} (see discussion in  Supplemental Material Sec.1\cite{SM}, which also includes Refs. \cite{MJ1983,Yan2019,ZSL2022,cao2024,HHY2024,MOO1989,MOO1985}). To ensure the high quality of 2H-NbS\textsubscript{2} flakes, we have developed a modified dry-transfer technique\cite{ZYC2022} and encapsulated the flakes with hexagonal boron nitride. As a result, the flake typically exhibits a residual resistivity ratio (RRR) exceeding 20 and a sharp superconducting transition as shown in Fig.\ref{fig2}(a). The critical temperature T\textsubscript{c} of flakes with thickness spanning from 45 nm down to 3.5 nm are summarized in Fig.\ref{fig2}(b). The raw data are enclosed in Supplemental Material Sec.2\cite{SM}. It is clear that T\textsubscript{c} drops monotonically with decreasing thickness, and can be fitted by\cite{XWZ2016,Y2023},     
\begin{equation}
	T_{cN}=T_{c1}+2\frac{J}{\alpha_0}cos(\frac{\pi}{N+1})
	\label{eq1}
\end{equation}
where $T_{c1}$ denotes critical temperature of a monolayer, $J$ quantifies interlayer Josephson coupling, $\alpha_0$ is related to density of state of a monolayer. From the fitting, it is concluded that $T_{c1}=1.30$ K and $\frac{J}{\alpha_0}=2.21$ K, and these parameters would be later on fed to the fitting of orbital FFLO.

The signature of the orbital FFLO state is encoded in the upper critical field. The temperature dependence of the upper critical field, defined as the field value at which RH reaches 50$\%$ of resistance of normal state (see Supplemental Material Sec.3\cite{SM}), in a 19 nm flake is recorded in Fig.\ref{fig3}(a). The raw data for flakes with thickness spanning from 9 nm to 45 nm are appended in Supplemental Material Sec.4\cite{SM}. The out-of-plane upper critical field $B_{c2,\perp}$ follows 3D Ginzburg-Landau (GL) relation. On the other hand,  the in-plane upper critical field $B_{c2,||}$ is initially in good agreement with 2D GL model in high-temperature regime. However, a noticeable enhancement in $B_{c2,||}$ occurs once the temperature drops below $\sim$ 5.2 K (0.89$T_{c0}$), denoted as $T^{\ast}$. The evolution of $B_{c2,||}$ in the entire temperature range can be satisfactorily reproduced by a orbital FFLO model which treats the system effectively as weakly coupled bilayer\cite{Y2023,L2017}. The enhancement of $B_{c2,||}$ marks the formation of orbital FFLO state with layer-dependent finite-momentum pairing below $T^{\ast}$ whereas trivial $B_{c2,||}$ behavior arising from zero-momentum pairing dictates high-temperature regime.  $T^{\ast}$ together with the upper critical field at this temperature, marked as $B^{\ast}$, defines a tricritical point\cite{Y2023, WZY2023} where orbital FFLO state, zero-momentum pairing superconducting state, and normal state merge. Note that the onset of the enhancement occurs at a magnetic field of 0.3$B_{p}$, much smaller than the field to stabilize a Zeeman FFLO state.

\begin{figure*}[h]
    \centering
\includegraphics[width=0.63\textwidth]{F3.jpg}
\caption{\textbf{Orbtial FFLO state as a function of flake thickness}. \textbf{a.} Thickness dependencies of \(B_{c2,\parallel}/B_p\) as a function of \(T/T_{c0}\). Traces are offset vertically for clarity. Enlarged version of results for each thickness can be found in Supplemental Sec.4\cite{SM}. We stress that the linear-like $B_{c2,||}-T$ relation in thicker flakes does not corresponds to the orbital limit of zero-momentum pairing 3D superconductivity. \textbf{b\&c.} Polar angle dependence of $B_{c2,||}$ in 25 nm and 30 nm flake, respectively. \textbf{d.} Orbital FFLO and Klemm-Luther-Beasley (KLB) analysis of $B_{c2,||}$ in 25 nm flake. The detail can be found in Supplemental Material Sec.4\cite{SM}. \textbf{e.} Polar angle dependence of $B_{c2,||}$ in 45 nm flake. }
\label{fig4}
\end{figure*}
 
Orbital FFLO state is featured by layer-dependent stripe-like oscillation in order parameter and therefore breaks both translational and rotational symmetry. The direct consequence of translational symmetry breaking is the sensitivity against the out-of-plane magnetic field component, following the general trend of inhomogeneous superconducting state in layered superconductors\cite{B1974,WZY2023}. In the experiment, we validate this feature by recording $B_{c2,||}$ with the magnetic field gradually tilting away from the in-plane direction. At T=5.7 K $> T^{\ast}$, the polar angle dependence of $B_{c2,||}$ takes a 2D-GL form as shown in Fig.\ref{fig3}(b). On the other hand,  the polar angle dependence has to be piecewise fitted by two 2D-GL components at T=4 K $< T^{\ast}$. The sharp component arising from spatial modulation in order parameter\cite{WZY2023} is responsible for $|\theta| < 1^{\circ}$. The broad component associated with uniform order parameter, restored by vortex motion, takes over at large $\theta$\cite{B1974,WZY2023}. Breaking rotational symmetry due to the formation of orbital FFLO state should install six-fold in-plane anisotropy against the magnetic field in TMDC flake\cite{WZY2023}. In our experiment, we have observed a dominating two-fold anisotropy arising from the vortex dynamics \cite{YAG2017} along the entire $B_{c}-T$ phase boundary. An incipient signature of six-fold anisotropy may emerge in the large field regime after subtracting the two-fold anisotropy in our study (see Supplemental Material Sec.5\cite{SM}).  

The observed $B_{c2,||}-T$ relation along with polar angle dependence of $B_{c2,||}$ can exclude other mechanisms that potentially lead to exotic $B_{c2,||}$ behavior, such as dimensional crossover\cite{ZDK2023,Roch2019}, two-band superconductivity\cite{BLH2022},  sample warping \cite{HB2018,KFR2008,SMG2022,JOI2015}. Details can be found in Supplemental Material Sec.6\cite{SM}.

\section*{Discussion}

Observation of orbital FFLO in 2H-NbS\textsubscript{2} has far-reaching impact. First, the qualitative similarity of orbital FFLO in 2H-NbS\textsubscript{2}, 2H-NbSe\textsubscript{2}\cite{WZY2023} and Li-intercalated  MoS\textsubscript{2}\cite{ZDK2023} encourages the search of orbital-FFLO state in other materials. Second and perhaps more interesting, the observation of orbital FFLO state in 2H-NbS\textsubscript{2} flake with intermediate thickness together with the fact that Ising superconductivity is a general feature of few-layer TMDC\cite{DSG2018,XWZ2016,JOI2015} and  Zeeman FFLO state is reported in NbS\textsubscript{2} bulk\cite{CLN2021,CCC2022} but not firmly established in other TMDC materials, suggesting there may be an intriguing evolution of Ising superconductivity $\longrightarrow$ orbital FFLO $\longrightarrow$ Zeeman FFLO with increasing flake thickness in 2H-NbS\textsubscript{2} as a consequence of interplay between Ising SOC and interlayer coupling.

\begin{figure}[h]
    \centering
\includegraphics[width=0.35\textwidth]{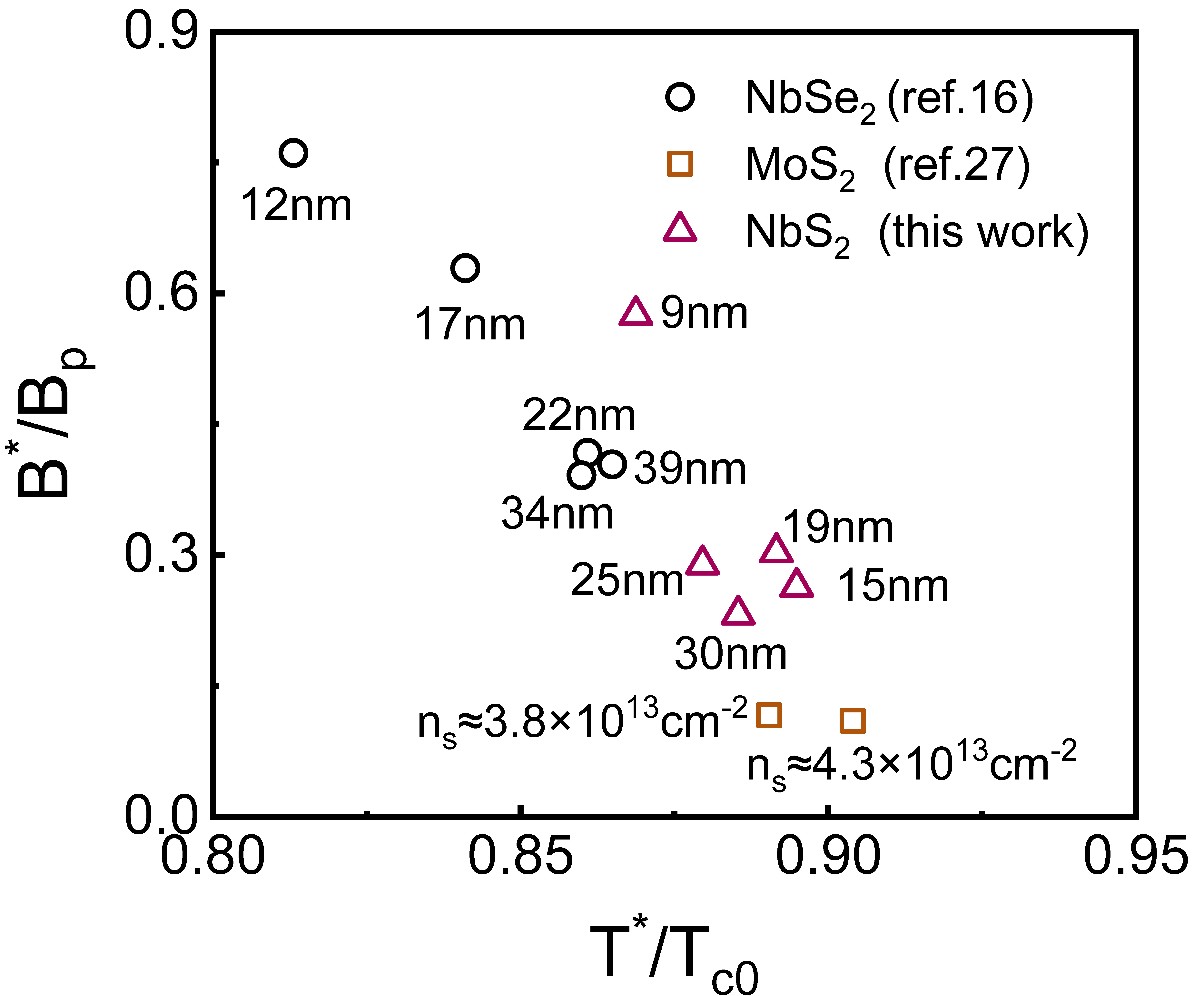}
\caption{\textbf{Summary of tricritical points of orbital FFLO state in different TMDC materials.} Data for NbSe\textsubscript{2} flakes come from ref.\cite{WZY2023} while that for Li-intercalated  MoS\textsubscript{2} bilayer are extracted from ref\cite{ZDK2023}. Note that the temperature axis only covers a narrow range between 0.80$T_c$ to 0.95$T_c$, so that the data appear to be more scattered along this axis.    }
\label{fig5}
\end{figure}

To shed light on the possible evolution of unconventional superconductivity in NbS\textsubscript{2}, we summarize $B_{c2,||}$ as a function of flake thickness in Fig.\ref{fig4}(a). $B_{c2,||}$ agrees well with the orbital FFLO model in thin flake of 19, 15 and 9 nm. It is helpful to mark that a recent work demonstrates that orbital FFLO state is absent in NbS\textsubscript{2} flake of 6 nm \cite{POV2024}. $T^{\ast}$ of the tricritical point drifts from 0.89$T_{c0}$ to 0.87$T_{c0}$ with decreasing thickness. Meanwhile, $B^{\ast}$ sees a jump from 0.30$B_p$ (19 nm) to 0.58$B_p$ (9 nm) while $B_p$ itself only changes by 10$\%$. At first glance, a reduction in $T^{\ast}$ accompanied by an enhancement in $B^{\ast}$ can be explained by an increment in the strength of interlayer Josephson coupling $J$\cite{ZDK2023} with decreasing thickness, however, this is against the fact that $J$ weakens in thinner flake\cite{XWZ2016}. Instead, the doubling of $B^{\ast}$ is more likely marking the onset of the interplay between orbital FFLO and Ising superconductivity, especially considering Ising SOC already has pronounced impact in 6 nm flake\cite{POV2024}. On the other hand, $B_{c2,||}$ of thicker flakes changes almost linearly against temperature. However, it is noteworthy that $B_{c2,||}$ is enhanced compared to the best linear fitting at a temperature below $T_d$=0.88$T_{c0}$ and $T_d$=0.89$T_{c0}$ for 25 nm and 30 nm flake. We emphasize that the deviation is always present regardless of the detail of the linear fitting, rather it has an intrinsic origin as uncovered by the polar angle dependence of $B_{c2,||}$. It requires dual-2D GL components to capture polar angle dependence below $T_d$ while one 2D component is sufficient for temperature above $T_d$ in 25 nm flake as shown in Fig.\ref{fig4}(b), qualitatively similar to the results in thinner flakes. 30 nm flake follows a similar behavior, see Supplemental Material Sec.4\cite{SM}. Inspired by this, we speculate it is perhaps necessary to attribute the two linear-like segments in $B_{c2,||}$-T to different mechanisms: The low-temperature segment, accompanied by dual-2D GL components, still arises from orbital FFLO state. The high-temperature segment does not correspond to the orbital-limit of zero-momentum pairing of 3D superconductivity, otherwise the polar angle dependence should agree with 3D GL model. Instead, it is a result of vortex depinning\cite{ZDK2023}. Key properties extracted from the vortex depinning effect in high temperature regime, described by Klemm-Luther-Beasley theory, such as the anisotropy in the in-plane and out-of-plane coherence length and spin life time, are in reasonable agreement with that expected for 2H-NbS\textsubscript{2} (see Supplemental Material Sec.4\cite{SM}). Hence, ($T_d$, $B_d$) marks the tricritical point in 25 and 30 nm flake. It is necessary to note that piecewise linear-like $B_{c2,||}-T$ relation has been reported in thick NbSe\textsubscript{2} flakes\cite{WZY2023} though the polar angle dependence is not presented, we assume that a similar orbital FFLO+KLB analysis can be carried out to test our hypothesis. Likewise, the nature of the linear $B_{c2,||}-T$ relation in 45 nm flake should also be carefully examined with the help of the polar angle dependence of $B_{c2,||}$. It is surprising that the polar angle dependence of $B_{c2,||}$ at both 0.95$T_{c0}$ and 0.75$T_{c0}$ comprise two 2D-GL components as enclosed in Fig.\ref{fig4}(c). Therefore, the observed linear $B_{c2,||}-T$ relation again does not correspond to orbital-limit of zero-momentum pairing 3D superconductivity. This observation is consistent with a recent experiment in $4H_{b}-TaS_{2}$ bulk\cite{FHS2024}, where it is suggested that the upper critical field in the entire temperature range is dominated by the orbital FFLO state. We leave a more thorough investigation on the nature of superconducting state in 45 nm flake to future work. 

Equipped with the thickness dependence data, it is helpful to draw a quantitative comparison between orbital FFLO states in different TMDC materials to highlight the key properties that may determine the formation of orbital FFLO state. We summarize the tricritical points of NbS\textsubscript{2}, NbSe\textsubscript{2}\cite{WZY2023} and Li-intercalated  MoS\textsubscript{2}\cite{ZDK2023} in Fig.\ref{fig5}. It is interesting to note that $T^{\ast}/T_{c0}$ distributes within a narrow window while $B^{\ast}/B_{p}$ spans in a large range. Since NbS\textsubscript{2} and NbSe\textsubscript{2} are operated in the absence of intercalation whereas the formation of orbital FFLO in MoS\textsubscript{2} depends on intercalation, it is expedient to focus on the difference between NbS\textsubscript{2} and NbSe\textsubscript{2} in this work. Then, it is recognized that $B^{\ast}/B_{p}$ is systemically smaller and $T^{\ast}/T_{c0}$ is consistently larger in NbS\textsubscript{2} flake compared to its NbSe\textsubscript{2} counterpart with similar thickness. Within the framework of the effective bilayer model of orbital FFLO state, $B^{\ast}\propto\sqrt{J}$ for TMDC flakes with similar flake thickness\cite{Y2023}, with $J$ denotes interlayer Josephson coupling. Hence the results immediately imply that $J$ should be smaller in NbS\textsubscript{2} than that in NbSe\textsubscript{2}, naturally arising from the fact out-of-plane aniostropy in NbS\textsubscript{2} is considerably larger in comparison with NbSe\textsubscript{2} and will suppress $J$ \cite{CCC2022}. Interestingly, the weak $J$  in NbS\textsubscript{2} is also vital for the stabilization of Zeeman FFLO in NbS\textsubscript{2} bulk, while the stronger $J$ in NbSe\textsubscript{2} bulk either prefers zero-momentum pairing\cite{CCC2022} or orbital FFLO state\cite{CTK2024}.

\section*{Conclusion}
We have observed orbital FFLO state with finite-momentum pairing in 2H-NbS\textsubscript{2} flakes of intermediate thickness, signified by the enhancement of in-plane upper critical field $B_{c2,||}$ and a nontrivial polar angle dependence of $B_{c2,||}$ constituted by two 2D-GL components. Orbital FFLO state, zero-momentum pairing superconducting state, and normal state merge at a tricritical point ($T^\ast$,$B^\ast$). By studying the thickness dependent of the tricritical point, we find a hint of interplay between Ising superconductivity and orbital FFLO showing as a jump in $B^\ast$ accompanied by a small change in $T^\ast$ with decreasing thickness in thin flake. By comparing the results in NbS\textsubscript{2} and NbSe\textsubscript{2} flake, it reveals that the relatively weak interlayer Josephson coupling is key in activating orbital FFLO in NbS\textsubscript{2} flakes at less demanding condition compared to NbSe\textsubscript{2}. With the observation of orbital FFLO state, our results highlight that NbS\textsubscript{2} can be a versatile platform to explore Ising superconductivity in few-layer limit, orbital FFLO state in flakes with intermediate thickness and Zeeman FFLO state in bulk-limit.

\clearpage
\noindent \textbf{Methods}

\noindent\textbf{Device fabrication.} 
We fabricated Ti/Au bilayer electrodes (15 nm Ti/35 nm Au) on a 285-nm-thick SiO\textsubscript{2} substrate using photolithography and electron-beam evaporation. High-quality NbS\textsubscript{2} flakes were mechanically exfoliated from bulk single crystals and transferred onto the pre-patterned substrate via an optimized dry transfer method with polydimethylsiloxane (PDMS) in a nitrogen-filled glovebox (H\textsubscript{2}O <0.01 ppm, O\textsubscript{2} <0.01 ppm). To ensure sample integrity during characterization, the NbS\textsubscript{2} flakes were encapsulated with hexagonal boron nitride (h-BN) as a protective capping layer. The devices were mounted on a single-axis rotator with 0.01° angular precision for alignment. Electrical transport measurements were subsequently performed in a Physical Property Measurement System (PPMS) and an Oxford cryostat. The thickness of the NbS\textsubscript{2} flakes was  determined by atomic force microscopy (AFM).

\clearpage
\bibliography{ref}

\vspace*{10pt}
\noindent \textbf{Acknowledgements} 

\noindent We acknowledge the financial support from the National Natural Science Foundation of China (Grant No. 12204184, 12074134, 12174021, 12074133) and the Innovation Program for Quantum Science and Technology (Grant No. 2021ZD0302700).\\

\noindent \textbf{Author contributions} 

\noindent  C. Y. and S. W. conceived and designed the experiment. X. Z. fabricated the samples. X. Z. and C. Y. conducted experiments and analyzed the data. G. G., X. L. and N. Y. performed the theoretical calculations. C. Z., C. L., H. G. and Y. L.  assisted in the experiment setup. C.Y. wrote the manuscript with input of other authors. C.Y., X. L and S. W. supervised the project. All authors discussed the results and contributed to the manuscript.\\

\noindent \textbf{Competing interests} 

\noindent The authors declare no competing interests. \\

\noindent \textbf{Data availability}

\noindent The data that support the findings of this study are available from the corresponding author upon reasonable request. \\

\noindent \textbf{Code availability}

\noindent The code that support the findings of this study are available from the corresponding author upon reasonable request. \\

\clearpage

\end{document}